# Fiber Nonlinearity Mitigation via the Parzen Window Classifier for Dispersion Managed and Unmanaged Links

*Abdelkerim Amari, Xiang Lin\*, Octavia A. Dobre\*, Ramachandran Venkatesan\*, Alex Alvarado*
*Information and Communication Theory Lab, Signal Processing Systems Group, Department of Electrical Engineering, Eindhoven University of Technology, The Netherlands.*
*\* Faculty of Engineering and Applied Science, Memorial University, St. John's, Canada,*
*e-mail: a.amari@tue.nl*

**ABSTRACT**
Machine learning techniques have recently received significant attention as promising approaches to deal with the optical channel impairments, and in particular, the nonlinear effects. In this work, a machine learning-based classification technique, known as the Parzen window (PW) classifier, is applied to mitigate the nonlinear effects in the optical channel. The PW classifier is used as a detector with improved nonlinear decision boundaries more adapted to the nonlinear fiber channel. Performance improvement is observed when applying the PW in the context of dispersion managed and dispersion unmanaged systems.
**Keywords**: Machine learning, fiber nonlinearity, optical communications, Parzen window.

## 1. INTRODUCTION

Machine learning techniques have been investigated for different applications in optical fiber communication systems [1-3]. Among these applications, the mitigation of fiber nonlinearity, which is the major challenge limiting the information-carrying capacity of long haul optical fiber transmission, has attracted considerable interest.
Machine learning techniques provide multiple advantages in comparison with conventional channel model-based nonlinear compensation techniques, such as digital back-propagation, Volterra series-based nonlinear equalization, and first-order perturbation theory-based NLC [7]. Firstly, machine learning techniques can deal with both non-Gaussian deterministic and stochastic nonlinear effects. Secondly, they have the potential of a lower implementation complexity. Furthermore, no knowledge of the optical link parameters are required, which make them well suited for optical networks.

Several machine learning algorithms have been proposed to combat fiber nonlinearity, such as support vector machine, K-nearest neighbours, and supervised k-means clustering [9-12]. The principle of these techniques consists of creating nonlinear decision boundaries by taking into account the nonlinear distortions. In the presence of fiber nonlinearity, the noise distribution is no longer circularly symmetric. Therefore, the optimum symbol detection requires knowledge and full parameterization of the likelihood function [1].

In [13], we proposed a machine learning-based detection technique, known as the Parzen window (PW) classifier, to mitigate the nonlinear non-Gaussian noise. It has been shown in [13] that PW significantly improves the performance by combatting the nonlinear noise, due to both deterministic nonlinear effects and stochastic nonlinear signal- amplified spontaneous emission (ASE) noise interactions. Limited performance improvement was observed in [13] for a channel with Gaussian noise, such as the case of long-haul dispersion unmanaged (DUM) systems, which is well modeled by the Gaussian noise (GN) model [14].

In this paper, we evaluate PW detector in the context of dispersion managed (DM) and DUM systems. We consider low symbol rates, which results in highly nonlinear non-Gaussian noise. We also consider high symbol rates, in which the high accumulated dispersion and its interactions with fiber nonlinearity leads to a more circularly symmetric Gaussian noise. We show performance improvement in terms of Q-factor and transmission reach in comparison with minimum Euclidean distance (MED) detection.

## 2. PARZEN WINDOW-BASED DETECTION

The main idea of machine learning classification for signal detection, and in particular PW, is to design improved nonlinear decision boundaries more adapted to the nonlinear fiber channel, which take into account the fiber nonlinearity. PW mitigates the nonlinear non-Gaussian noise caused by deterministic nonlinear effects and also stochastic nonlinear signal- ASE noise interactions.

The principle of PW is introduced as follows. At the transmitter, a label is associated to each quadrature amplitude modulation (QAM) symbol; thus, $M$ classes exist in systems employing $M$-QAM. We use $\{L_m\}_{m\in[1,M]}$ to denote the labels for the $m$-th class. Then, $T$ training symbols $\{x_t\}_{1 \leq t \leq T}$ are generated along with $N$ testing symbols $\{x_n\}_{1 \leq n \leq N}$, and sent into the channel. At the receiver, the received training data $\{y_t\}_{1 \leq t \leq T}$ is employed to build the decision boundaries. More specifically, a likelihood metric of each received testing data $\{y_n\}_{1 \leq n \leq N}$ and all the received training data $\{y_t\}_{1 \leq t \leq T}$ is calculated to make the decision. Before defining the metric, two important factors of the PW are introduced: the window function $f$ and window size $R$. A kernelized window function is defined as

$$f_{n,t} = \begin{cases} \frac{1}{D(y_n, y_t)} & \text{if } D(y_n, y_t) < R, \\ 0 & \text{otherwise,} \end{cases} \quad (1)$$

where $D$ represents Euclidean distance. For this window function, the training data closest to the testing data has the highest impact on the decision. The window shape is considered as a circle, with radius $R$, because the data is distributed in 2 dimensions. The radius $R$ is optimised to achieve the best performance as in [13].

After the decision boundaries have been found, for each testing data $y_n$, the distances from each training data are collected. Then, the likelihood metric is defined as

$$S_{n,m} = \sum_{\substack{t=1 \\ x_t = L_m}}^{T} f_{n,t}, \quad m=1, 2, ..., M. \quad (2)$$

The maximum value determines the most likely label of the $n$-th symbol by

$$\hat{m}_n = \underset{m}{\arg\max}\ S_{n,m}, \quad m=1, 2, ..., M. \quad (3)$$

Finally, the label gives the corresponding QAM symbol.

Fig. 1 shows the PW-based decision boundaries and its corresponding decision regions designed using the training data in Fig. 1(a). By designing new decision boundaries taking into account for the fiber nonlinearity, PW can mitigate the non-Gaussian nonlinear noise and improve the system performance.

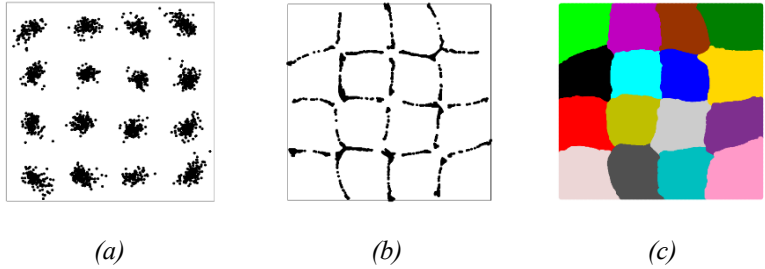

*(a)*      *(b)*      *(c)*

*Figure 1. Example of PW-based decision for DM system, used in Section. 3, in the nonlinear regime with 1200 km, -1 dBm input power, and 10 Gbaud: (a) received training data; (b) decision boundaries; (c) decision regions.*

## 3. SIMULATION SETUP AND RESULTS

The performance of PW-based detector is evaluated in comparison with MED detector in the context of 16-QAM dual-polarization single-channel transmission system. We consider DM with full-chromatic dispersion (CD) compensation, and DUM systems. The simulation setup is depicted in Fig. 2.

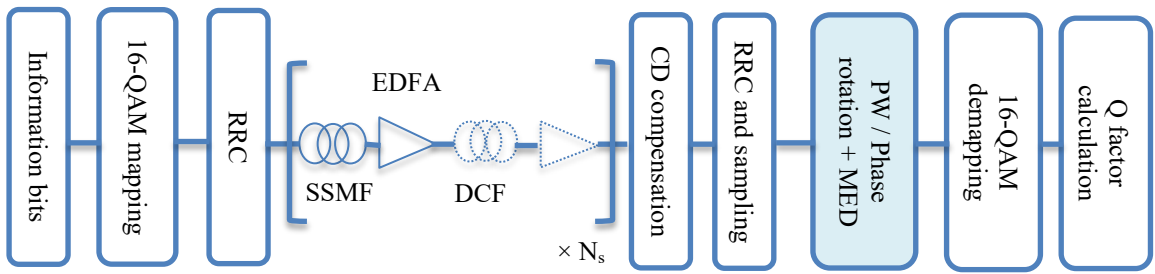

*Figure 2. Simulation setup. Dashed lines are considered only in DM system.*

At the transmitter side, the information bits are mapped into 16-QAM symbols. Then, a root-raised cosine (RRC) filter with 0.1 roll-off factor is applied for spectrum shaping. The transmission link consists of multi-span standard single mode fiber (SSMF) with an attenuation coefficient α= 0.2 dB / km, a dispersion parameter Ð = 16 ps / (nm×km), and a nonlinear coefficient γ= 1.4 / (W×km). After each span with 80 km length, the signal is amplified using an erbium-doped fiber amplifier (EDFA) with a 5.5 dB noise figure and 16 dB gain. Concerning the DM system, a dispersion-compensated fiber and an EDFA are added for each span. At the receiver side, chromatic dispersion (CD) compensation is performed before applying matched filter and downsampling. When applying the PW detector, the phase rotation compensation is not required because the detection relies on the labeled training symbols [13]. In the case of MED detection, the phase rotation is compensated using training sequence. Finally 16-QAM demapping is performed before Q-factor calculation, which is done as in [4]. 2000 symbols are used as training symbols for PW detection.

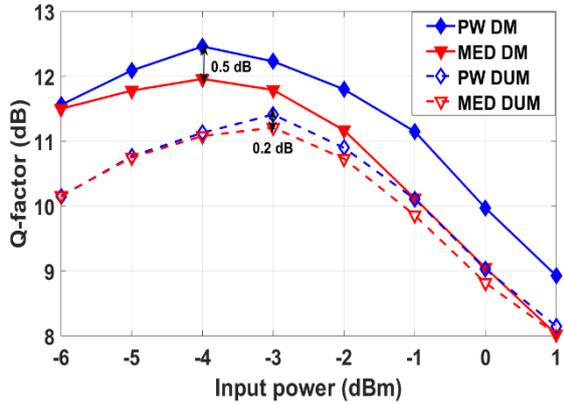
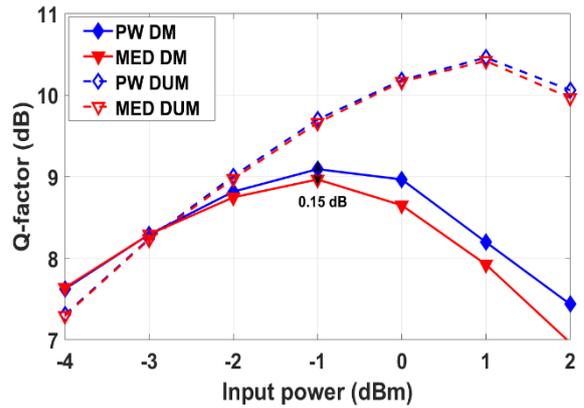

Figure 3. Q-factor vs. input power for 10 Gbaud and 1200 km.

Figure 4. Q-factor vs. input power for 45 Gbaud and 1200 km.

In Fig. 3, the Q-factor is plotted as a function of the input power for 10 Gbaud symbol rate and 1200 km for DM and DUM systems. For DM system, PW outperforms the MED-based detector, and the gain in terms of Q-factor at optimal input power (-4 dBm) is about 0.5 dB. PW provides about 1 dB improvement in the nonlinear regime at high input power, while similar performance to the MED detection is observed in the linear regime (-6 dBm). This confirms that, by designing improved nonlinear decision boundaries, fiber nonlinearity can be mitigated, which leads to significant performance improvement. Concerning the DUM system, PW-based detector exhibits about 0.2 dB Q-factor gain in comparison with MED detector at optimal input power (-3 dBm).

In Fig. 4, the symbol rate is increased to 45 Gbaud. It is observed that the gain of PW, at optimal input power (-1 dBm) is reduced to about 0.15 dB in the DM system. In the DUM case, similar performances of PW and MED-based detectors are observed. These results can be explained by the fact that, for higher symbol rate, the dispersion effect increases, and its interaction with the fiber nonlinearity results in a Gaussian-like noise. In case of channel with Gaussian noise, the MED detector provides good performance and its corresponding decision regions are well suited for signal detection in such systems. The designed nonlinear decision boundaries based on PW are similar to those of the MED.

The received constellations, shown in Fig. 5, confirm this analysis. For DUM system the constellations for 45 Gbaud (Fig. 5(b)) is more Gaussian than constellations for 10 Gbaud (Fig. 5(a)). Similar observation is obtained for DM system. Therefore, the performance improvement that PW provides, in comparison with MED, increases when the nonlinear noise becomes more and more non-Gaussian.

We also evaluated the gain in terms of transmission reach that PW provides in comparison with MED detector for 10 Gbaud DM system. As shown in Fig. 6, PW increases the reach by about 160 km in comparison with MED, at a Q-factor of 10 dB.

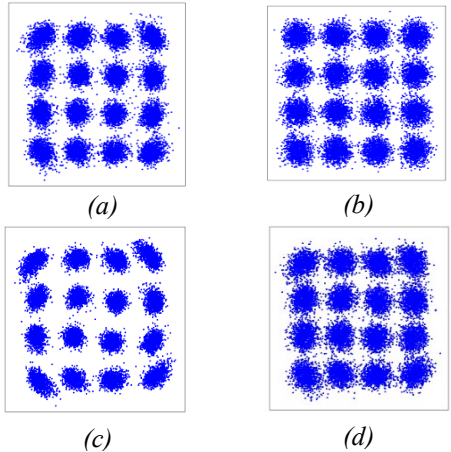

Figure 4. Constellation diagrams at -1 dBm: (a) 10 Gbaud DUM, (b) 45 Gbaud DUM, (c) 10 Gbaud DM, (d) 45 Gbaud DM.

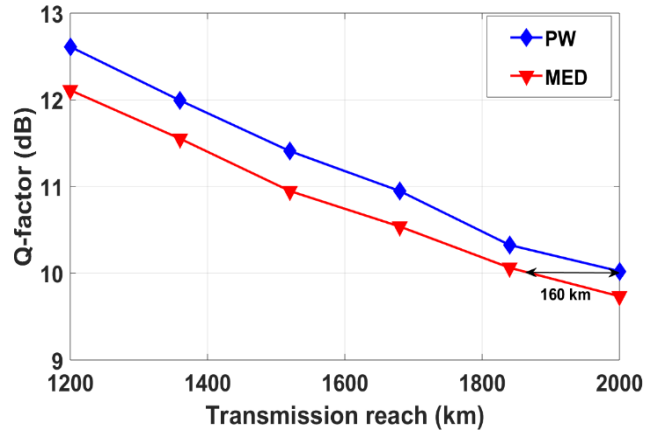

Figure 5. Q-factor vs. transmission reach for 10 Gbaud DM at optimal input power (-4 dBm).

## CONCLUSIONS

We have shown that Parzen window (PW) detector mitigates the non-Gaussian fiber nonlinear effects by designing nonlinear decision boundaries. PW improves the performance in comparison with minimum distance-based detection in dispersion managed and low symbol rate dispersion unmanaged systems. A significant Q-factor and transmission reach increase is observed for low symbol rate dispersion managed system.

## ACKNOWLEDGEMENTS

This work was supported by the Netherlands Organization for Scientific Research (NWO) via the VIDI Grant ICONIC (project number 15685).

## REFERENCES

[1] F. Musumeci, C. Rottondi, and M. Tornatore, "A survey on application of machine learning techniques in optical networks," arXiv: 1803.07976 [cs. NI], Apr. 2018.
[2] D. Rafique and L. Velasco, "Machine learning for network automation: Overview, architecture, and applications [invited tutorial]," *J. Opt. Commun. Netw.*, vol. 10, no. 10, pp. D126–D143, Oct. 2018.
[3] X. Lin, *et al*., "Joint modulation classification and OSNR estimation enabled by support vector machine," *IEEE Photon. Technol. Lett.*, vol.30, no. 24, pp. 2127-2130, Dec. 2018.
[4] A. Amari *et al.*, "A survey on fiber nonlinearity compensation for 400 Gb/s and beyond optical communication systems," *IEEE Commun. Surveys Tuts.*, vol. 19, no. 4, pp. 3097-3113, 4th Quart., 2017.
[5] E. Ip *et al*., "Compensation of dispersion and nonlinear impairments using digital back propagation," *J. Lightwave Technol.*, vol. 26, no. 20, pp.3416-3425, 2008.
[6] A. Amari, P. Ciblat, Y.Jaouen, "Fifth-order Volterra series based nonlinear equalizer for long-haul high data rate optical fiber communications," in *Proc. ACSSC*, Nov. 2014, pp. 13671371.
[7] X. Liang *et al*., "Digital compensation of cross-phase modulation distortions using perturbation technique for dispersion-managed fiber-optic systems," *Opt. Express*, vol. 22, no. 17, pp. 20 63420 645, Aug. 2014.
[8] A. Amari, P. Ciblat, Y. Jaouën, "Inter-subcarrier nonlinear interference canceler for long-haul Nyquist-WDM transmission", *IEEE Photon. Technol. Lett.,* vol. 28, no. 23, pp. 2760-2763, Dec. 2016.
[9] D. Zibar *et al*., "Machine learning techniques applied to system characterization and equalization," in *Proc. Opt. Fiber Commun. Conf.*, Mar. 2016, paper Tu3K.1.
[10] M. Li *et al*., "Nonparameter nonlinear phase noise mitigation by using M-ary support vector machine for coherent optical systems," *IEEE Photonics J.*, vol. 5, no. 6, Dec. 2013.
[11] D. Wang *et al*., "Nonlinearity mitigation using a machine learning detector based on k-nearest neighbors," *IEEE Photon. Technol. Lett.*, vol. 28, no. 19, pp. 2102-2105, Oct. 2016.
[12] J. Zhang *et al*., "K-means-clustering-based fiber nonlinearity equalization techniques for 64-QAM coherent optical communication system," *Opt. Express*, vol. 25, no. 22, pp. 27570-27580, Oct. 2017.
[13] A. Amari *et al.*, "A machine learning-based detection technique for optical fiber nonlinearity mitigation," *IEEE Photon. Technol. Lett.*, vol.31, no. 08, pp. 627-630, Apr. 2019.
[14] P. Poggiolini, "The GN model of non-linear propagation in uncompensated coherent optical systems," J. Lightw. Technol., vol. 30, no. 24, pp. 3857-3879, Dec. 2012.